\documentclass[10pt,aps,prl,amsmath,amssymb,superscriptaddress,floatfix,nofootinbib,notitlepage,reprint]{revtex4-1}
\usepackage{amsmath,amssymb}
\usepackage{graphicx,color}
\usepackage[colorlinks=true,hyperfootnotes=false]{hyperref}
\usepackage{epstopdf}
\usepackage{txfonts}

\begin{document}
\title{\boldmath $Z_c(3900)$: what has been really seen?}
\author{Miguel Albaladejo}
%\email{Miguel.Albaladejo@ific.uv.es}
\affiliation{Instituto de F\'isica Corpuscular (IFIC),
             Centro Mixto CSIC-Universidad de Valencia,
             Institutos de Investigaci\'on de Paterna,
             Aptd. 22085, E-46071 Valencia, Spain}
\author{Feng-Kun Guo}
%\email{fkguo@hiskp.uni-bonn.de}
\affiliation{State Key Laboratory of Theoretical Physics,
            Institute of Theoretical Physics, Chinese Academy of Sciences,
            Beijing 100190, China}
\affiliation{Helmholtz-Institut f\"ur Strahlen- und
             Kernphysik and Bethe Center for Theoretical Physics, \\
             Universit\"at Bonn,  D-53115 Bonn, Germany}
\author{Carlos Hidalgo-Duque}
%\email{carloshd@ific.uv.es}
\affiliation{Instituto de F\'isica Corpuscular (IFIC),
             Centro Mixto CSIC-Universidad de Valencia,
             Institutos de Investigaci\'on de Paterna,
             Aptd. 22085, E-46071 Valencia, Spain}
\author{Juan Nieves}
%\email{jmnieves@ific.uv.es}
\affiliation{Instituto de F\'isica Corpuscular (IFIC),
             Centro Mixto CSIC-Universidad de Valencia,
             Institutos de Investigaci\'on de Paterna,
             Aptd. 22085, E-46071 Valencia, Spain}

\begin{abstract}

The $Z^\pm_c(3900)/Z^\pm_c(3885)$ resonant structure has been experimentally
observed in the $Y(4260) \to J/\psi \pi\pi$ and $Y(4260) \to \bar{D}^\ast D \pi$
decays. This structure is intriguing since it is a prominent candidate of an
exotic hadron. Yet, its nature is unclear so far. In this work, we
simultaneously describe the $\bar{D}^\ast D$ and $J/\psi \pi$ invariant mass
distributions in which the $Z_c$ peak is seen using amplitudes with exact
unitarity.
Two different scenarios are statistically acceptable, where the origin of the
$Z_c$ state is different. They correspond to using energy dependent or
independent $\bar D^* D$ $S$-wave interaction. In the first one, the $Z_c$ peak
is due to a resonance with a mass around the $D\bar D^*$ threshold. In the second one, the $Z_c$ peak is produced by a virtual state which must have a
hadronic molecular nature. In both cases the two observations, $Z^\pm_c(3900)$
and $Z^\pm_c(3885)$, are shown to have the same common origin, and a
$\bar D^* D$ bound state solution is not allowed. Precise measurements of the line shapes around the $D\bar D^*$ threshold are
called for in order to understand the nature of this state.

\end{abstract}

\maketitle

% \section{Introduction}
The resonant-like structure $Z_c(3900)^{\pm}$ was first seen simultaneously by
the BESIII and Belle collaborations \cite{Ablikim:2013mio, Liu:2013dau} in the
$J/\psi \pi$ spectrum produced in the $e^+ e^- \to Y(4260) \to J/\psi\pi^+ \pi^-
$ reaction. An analysis \cite{Xiao:2013iha} based on CLEO-c data for the $e^+
e^- \to \psi(4160) \to  J/\psi\pi^+\pi^-$ reaction confirmed the presence of
this structure as well, although with a somewhat lower mass. Under a different name, $Z_c(3885)^{\pm}$, a similar structure, with
quantum numbers favored to be $J^P=1^+$, has also been reported by the BESIII
collaboration~\cite{Ablikim:2013xfr, Ablikim:2015swa} in the $\bar{D}^\ast D$
spectrum of $e^+ e^-\to \bar D^* D\pi$ at different $e^+e^-$ center-of-mass
(c.m.) energies [including the production of $Y(4260)$]. Because there is
a little difference in the central values of the masses and in particular the
widths of these two structures, whether they correspond to the same state is
still unknown. As will be shown in this Letter, the two structures have indeed
the same common origin. We generically denote it here
as $Z_c$. Evidence for a neutral partner of this structure was first reported in
Ref.~\cite{Xiao:2013iha}, and more recently in Ref.~\cite{Ablikim:2015tbp}.

If this resonant structure happens to be a real state as argued in
Ref.~\cite{Guo:2014iya}, it is one of the most interesting hadron resonances,
since it couples strongly to charmonium and yet it is charged, thus it is
something clearly distinct of a conventional $c\bar{c}$ state --- its minimal
constituent quark content should be four quarks, $c\bar{c} u \bar{d}$ (for
$Z_c^+$). A discussion of possible internal structures is given in
Ref.~\cite{Voloshin:2013dpa}. It has been interpreted as a molecular
$\bar{D}^\ast D$ state~\cite{Wang:2013cya,Guo:2013sya,int:molecules}, as a
tetraquark of various configurations~\cite{int:tetraquarks}
% , as a hybrid \cite{Braaten:2013boa}
or as a simple kinematical effect~\cite{int:cusps}, although this possibility
has been ruled out in Ref.~\cite{Guo:2014iya}. Distinct consequences of some of
these different models have been discussed in Ref.~\cite{Cleven:2015era}. It has
been also searched for in lattice QCD though with negative results so
far~\cite{int:LQCD}.

Being a candidate for an explicitly exotic hadron, the $Z_c(3900)$ definitely
deserves a detailed and careful study. Indeed, the last years have witnessed an
intense theoretical activity aiming at understanding the actual nature of this
state. What is still missing, however, is a simultaneous study of the two
reactions analysed by BESIII and mentioned above in which the $Z_c$ structure
has been seen.\footnote{Both reactions were considered in
Ref.~\cite{Guo:2014iya} and used to fix parameters at the one-loop level.
The purpose there is to show that the narrow near threshold states like the $Z_c$
cannot be simply kinematical effects. Despite  the importance, a detailed
global analysis of the data for both reactions using fully resummed and
unitarized amplitudes has not been done before.} The goal of this work is to perform such a study, and, from it, to extract
information about this seemingly resonant intriguing structure. We will first
settle a $\bar{D}^\ast D$, $J/\psi \pi$ coupled channel formalism, considering
that the $Z_c$ emerges from the $\bar{D}^\ast D$ interaction, and that its
coupling to $J/\psi \pi$ proceeds through the former intermediate state. The
resulting $T$-matrix will enter the calculation of the amplitudes for the
reactions $Y(4260) \to J/\psi \pi \pi, \bar{D}^\ast D \pi$. We will assume that
the $Y(4260)$ state is dominantly a $D_1(2420)\bar{D}+\text{c.c.}$ bound
state \cite{Wang:2013cya,Y4260molecule} and use the ideas of Ref.~\cite{Wang:2013cya}
to compute the relevant amplitudes.

% \section{Formalism}
Let us denote with 1 and 2 the $J/\psi\pi$ and $\bar{D}^\ast D$ channels,
respectively, with $I=1$ and $J^{PC}=1^{+-}$ (here and below, the $C$-parity
refers to the neutral member of the isospin triplet). The coupled-channel
$T$-matrix can be written as
\begin{equation}
T = (\mathbb{I}-V \cdot G)^{-1} \cdot V~,
\end{equation}
where $G$ is the loop function diagonal matrix, and the matrix elements of the
 potential read
\begin{equation}\label{eq:potential}
V_{ij} = 4 \sqrt{m_{i1} m_{i2}} \sqrt{m_{j1} m_{j2}}\, e^{-q_i^2/\Lambda_i^2}
e^{-q_j^2/\Lambda^2} C_{ij}~,
\end{equation}
%
%In Eq.~\eqref{eq:potential}
where $m_{i\,n}$ is the mass of the the $n$th particle in
the channel $i$, and the mass factors are included to account for the
non-relativistic normalization of the heavy meson fields.
The $J/\psi \pi \to J/\psi \pi$ interaction strength is known to be
tiny~\cite{Yokokawa:2006td,Liu:2012dv}, and we neglect the direct coupling of
this channel, $C_{11}=0$. Such a treatment was also done in
Ref.~\cite{Hanhart:2015cua} in a coupled-channel analysis of the $Z_b$ states.
For the inelastic $\bar{D}^\ast D \to J/\psi \pi$ $S$-wave interaction,  we make
the simplest possible assumption, that amounts to take it to be a constant,
$C_{12} \equiv \widetilde{C}$.
In a momentum expansion, the lowest order contact potential for the
$\bar{D}^\ast D \to \bar{D}^\ast D$ transition is simply a constant as well, denoted by $C_{22} \equiv
C_{1Z}$~\cite{HQSSFormalism}. However, it can be shown that even with two
coupled channels, no resonance can be generated in the complex plane above threshold
with only constant potentials.
% (apart from the factorized energy dependence of the regulators).
To that end, we will also allow some energy dependence for the $V_{22}$
term, introducing a new parameter $b$, and writing
\begin{equation}\label{eq:C22_b}
C_{22}(E) = C_{1Z} + b\left(E-m_D - m_{D^\ast}\right)
\end{equation}
with $E$ the total c.m. energy.
The new term is of higher order in low-momentum expansion in comparison
with $C_{1Z}$.
The interactions
considered here need to be regularized in some way, and hence we employ a
standard gaussian regulator~\cite{Epelbaum:2008ga}, $e^{-q_i^2/\Lambda_i^2}$,
where the c.m.
momentum squared of the channel $i$ is denoted by ${q_i}^2$. We adopt a relativistic
(non-relativistic) definition of the latter for the $i=1$ ($i=2$) channel, {\it
i.e.}, ${q_1}^2 = \lambda(E^2,m_{J/\psi}^2,m_\pi^2)/(4E^2)$ and ${q_2}^2 =
2\mu(E-m_D-m_{D^\ast})$, being $\mu$ the reduced mass of the $\bar{D}^\ast D$
system. Since the interaction for this channel is derived from a
non-relativistic field theory, we take cutoff values $\Lambda_2 = 0.5 - 1\
\text{GeV}$~\cite{HQSSFormalism}. At the $Z_c$ energy, the c.m. momentum of the
$J/\psi \pi$ channel is $q_1 \simeq 0.7\ \text{GeV}$, and hence we use a
different cutoff for it. For definiteness, we set $\Lambda_1 = 1.5\ \text{GeV}$,
although the specific value is not very relevant as we have checked since
changes in the cut-off can be reabsorbed in the strength of the transition
potential controlled by the undertermined $C_{12}$ low energy constant. With
this convention for the regulator, the loop functions in the matrix $G$ read
\begin{align}
G_1(E) & = \int \frac{l^2 dl}{4\pi^2} \frac{\omega_1+\omega_2}{\omega_1 \omega_2} \frac{e^{-(l^2-q_1^2)/\Lambda_1^2}}{E^2-(\omega_1 + \omega_2)^2+i \epsilon}~,\label{eq:G1}\\
G_2(E) & = \frac{1}{m_D+m_{D^\ast}} \int \frac{l^2 dl}{4\pi^2}\frac{e^{-(l^2-q_2^2)/\Lambda_2^2}}{q_2^2 - l^2 + i\epsilon}~,\label{eq:G2}
\end{align}
with $\omega_n = \sqrt{l^2+m_{1n}^2}$. The $D\bar{D}^\ast$ channel loop function $G_2$ is computed in the non-relativistic approximation.

\begin{figure}
\centering\includegraphics[width=0.50\textwidth,keepaspectratio]{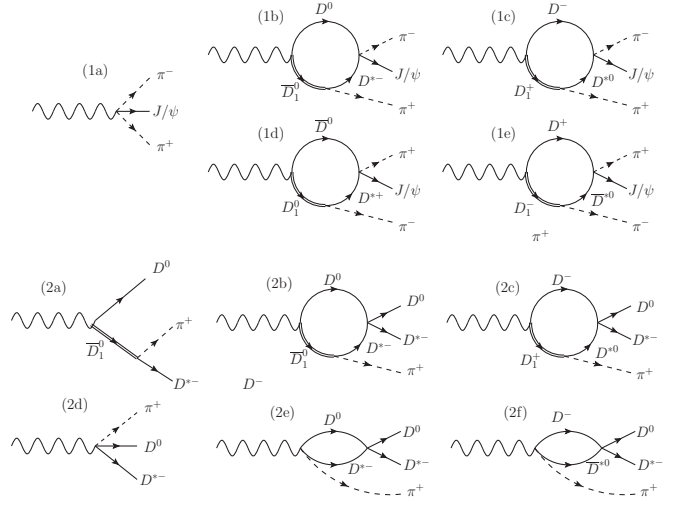}
\caption{Diagrams contributing to the $Y(4260)$ (wavy line) decays to $\bar{D}^\ast D\pi$ and $J/\psi\pi\pi$.\label{fig:diagrams}}
\end{figure}

For the $e^+e^-$ annihilations at the $Y(4260)$ mass, both BESIII and Belle have
reported the $Z_c$ structure in the $J/\psi \pi$ final state
\cite{Ablikim:2013mio,Liu:2013dau}, but only BESIII provides data for the
$\bar{D}^\ast D$ channel~\cite{Ablikim:2013xfr,Ablikim:2015swa}. Hence, for
consistency, we will only study the BESIII data. In particular, we will consider
the most recent double-$D$-tag data of Ref.~\cite{Ablikim:2015swa}, in
which the $D^\ast$ is reconstructed from several decay modes, whereas in
Ref.~\cite{Ablikim:2013xfr} the presence of the $D^\ast$ is only inferred from
energy conservation. Hence, in the former data the background in the higher
energy $D^\ast D$ invariant mass regions is much reduced. For definiteness, we
will consider the reported spectra of the $D^{\ast-}D^0$ and $J/\psi \pi^-$
final states, and set $m_{D^\ast} = m_{D^{\ast-}}$, $m_{D} = m_{D^0}$, and
$m_\pi = m_{\pi^{\pm}}$.
This implicitly assumes that isospin breaking effects are neglected. These data
are taken at a c.m. energy equal to the nominal $Y(4260)$ mass, so the decays to
$\pi(J/\psi \pi,  \bar{D}^\ast D)$ proceed mainly through the formation of this
resonance. The mechanisms for the $Y(4260)$ decays are shown in
Fig.~\ref{fig:diagrams}. The coupling $Y D_1 D$, whose value is not
important here to describe the lines shapes, is taken from
Ref.~\cite{Wang:2013cya}, where the $Y(4260)$ is considered to be
dominantly a $\bar{D} D_1 + \text{c.c.}$ bound state. The subsequent $D_1
D^\ast \pi$ coupling can also be found there.

We denote $\mathcal{M}_1$ ($\mathcal{M}_2$) to the amplitude for the $Y \to
J/\psi \pi^+ \pi^-$ ($Y \to D^{\ast-} D^0 \pi^+$) decay, and $s$ and $t$,
respectively, to the invariant masses squared of $J/\psi \pi^-$ and $J/\psi
\pi^+$ ($D^{\ast-}D^0$ and $D^{\ast-}\pi^+$) in the first (second) decay. Up to
some common irrelevant constant, both amplitudes can be written (after the
appropriate sum and average over polarizations) as:
\begin{align}
&  \left\lvert \overline{\mathcal{M}_1}(s,t) \right\rvert^2 = \left\lvert \tau(s) \right\rvert^2 q^4_\pi(s) + \left\lvert \tau(t) \right\rvert^2 q^4_\pi(t) \nonumber \\
& + \frac{3\cos^2\theta-1}{4} \big(\tau(s)\tau(t)^\ast + \tau(s)^\ast\tau(t) \big) q^2_\pi(s) q^2_\pi(t)~, \\
& \tau(s) = \sqrt{2} I_3(s) T_{12}(s) + \alpha~, \label{eq:tau}\\
&  \left\lvert \overline{\mathcal{M}_2}(s,t) \right\rvert^2 = \left\lvert
\frac{1}{t-m_{D_1}^2} +  I_3(s) T_{22}(s)\right\rvert^2 q^4_\pi(s) \nonumber \\
& + \left\lvert \beta \big( 1 + T_{22}(s) G_{2}(s) \big)\right\rvert^2~, \label{eq:M2}
\end{align}
where $q^2_\pi(s) = \lambda(M_Y^2,s,m_\pi^2)/(4M_Y^2)$, and $\theta$ denotes the relative angle between the two pions in the $Y(4260)$
rest frame. Further, $I_3(s)$ is the scalar three-meson non-relativistic loop
function, for which details can be found in Ref.~\cite{Albaladejo:2015dsa}. One
first notes that $\overline{\mathcal{M}_1}(s,t)$ is symmetric under $s
\leftrightarrow t$. The term with $\alpha$ represents diagram (1a), and it acts
as a non-resonant background amplitude, added coherently to the rest of the
diagrams.
It has the same dependence on the external momenta and polarization vectors as that of
diagrams (1b)--(1e). The first term in $\overline{\mathcal{M}_1}(s,t)$ is the
amplitude of diagrams (1b)+(1c), the second term is the one from diagrams
(1d)+(1e), and the last one is their interference. In
$\overline{\mathcal{M}_2}$, the first summand of the first term corresponds to
diagram (2a) in Fig.~\ref{fig:diagrams}, whereas the second one, which
includes the $\bar{D}^\ast D$ final state interaction (FSI), is the contribution
from diagrams (2b)+(2c). Diagrams (2a)-(2c) proceed through the formation of
$D_1$, but we also consider some non-resonant $\bar{D}^\ast D \pi$ production by
means of diagram (2d). The $\bar{D}^\ast D$ rescattering effects in this last
diagram give rise, in turn, to diagrams (2e) and (2f). The term with $\beta$ in
Eq.~\eqref{eq:M2} represents these latter three diagrams. The parameters
$\alpha$ and $\beta$ in Eqs.~\eqref{eq:tau} and \eqref{eq:M2} are unknown. Note that the effect of $D_1$ width, $\Gamma_{D_1}=(25\pm6)$~MeV, is
negligible here since $m_{D_1}+m_D-\Gamma_{D_1}/2$ is well above
4.26~GeV.\footnote{Inclusion of the $D_1$ width into the calculation of
$\Gamma(Y(4260)\to\gamma X(3872))$ only leads to a change of about
3\%~\cite{Guo:2013nza}.}

The spectrum for both reactions can be obtained as a contribution from the
amplitudes ($\mathcal{A}_i$) plus a background ($\mathcal{B}_i$):
\begin{align}
\mathcal{N}_i(s) & = K_i  \left( \mathcal{A}_i(s) +  \mathcal{B}_i(s) \right)~,\\
\mathcal{A}_i(s) & = \int_{t_{i,-}}^{t_{i,+}}\!\!\!\! dt \left\lvert \overline{\mathcal{M}_i}(s,t) \right\rvert^2~,
\end{align}
where $t_{i,\pm}(s)$ are the limits of the $t$ Mandelstam variables for the
decay mode $i$. The two global constants $K_i$ could be related if the event selection efficiencies of the two spectra analyzed in this work were known. If the latter were roughly the same, then one would have $K_1 \simeq 5 K_2$ (due to the different bin sizes). If both parameters are considered free, a large correlation arises between $K_1$ and $\widetilde{C}$, since $K_1 \lvert \widetilde{C} \rvert^2$ basically determines the total strength of the event distribution $\mathcal{N}_1$. This is due to the fact that the influence of $\widetilde{C}$ in the shape of the $T$-matrix elements, and thus of the signal of $Z_c$ in the spectrum, is small. To obtain a reasonable estimate of this coupling constant, we consider a further experimental input from Ref.~\cite{Ablikim:2013xfr},
\begin{equation}
R_\text{exp} = \frac{\Gamma\left(Z_c(3885) \to D\bar{D}^\ast\right)}{\Gamma\left(Z_c(3900) \to J/\psi \pi \right)} = 6.2 \pm 1.1 \pm 2.7~,
\end{equation}
and estimate this ratio as
\begin{equation}
R_\text{th} = \frac{\int ds \mathcal{A}_2(s)}{\int ds \mathcal{A}_1(s)}~,
\end{equation}
that is, as the ratio of the background subtracted areas of each physical
spectrum around the $Z_c$ mass, namely in the range $\sqrt{s} = (3900 \pm 35)\
\text{MeV}$.

\begin{figure*}
\includegraphics[width=0.33\textwidth,keepaspectratio]{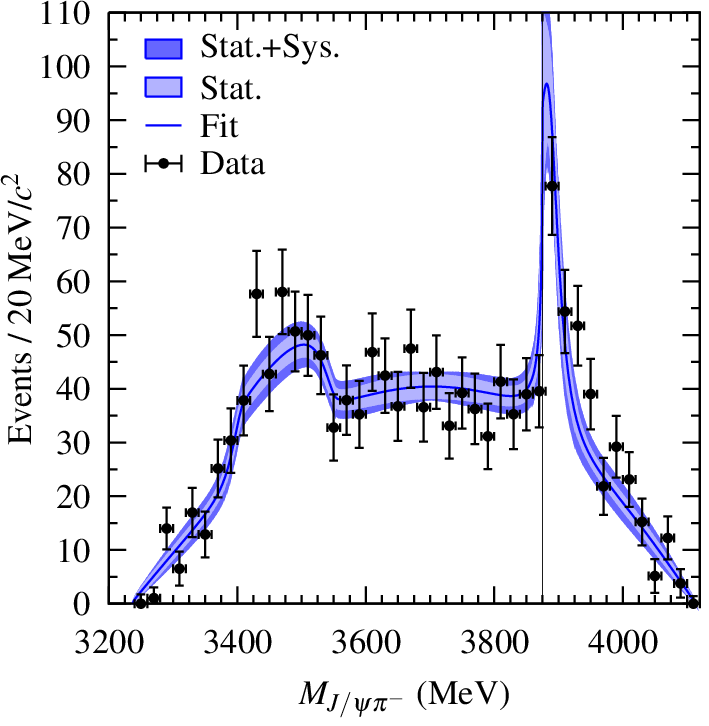}\hspace{2mm}%\\
\includegraphics[width=0.33\textwidth,keepaspectratio]{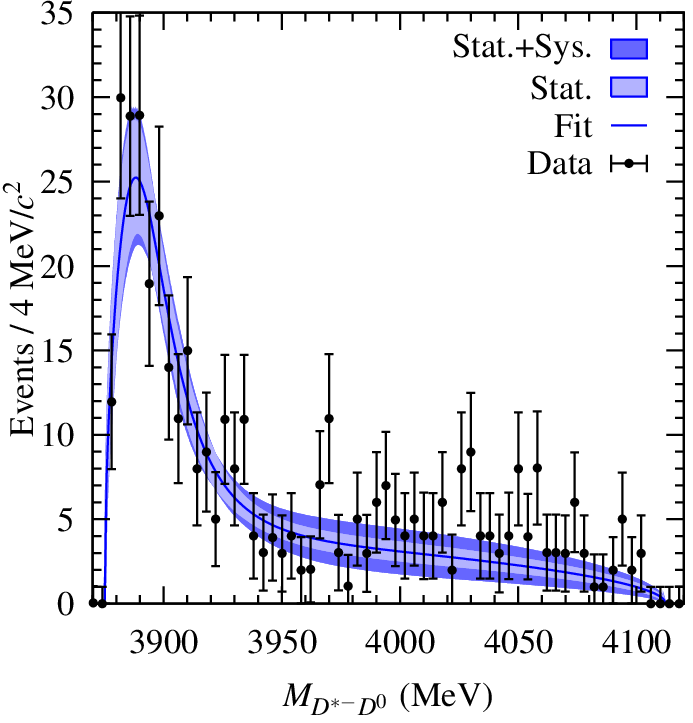}\\%\hspace{2mm}
\includegraphics[width=0.33\textwidth,keepaspectratio]{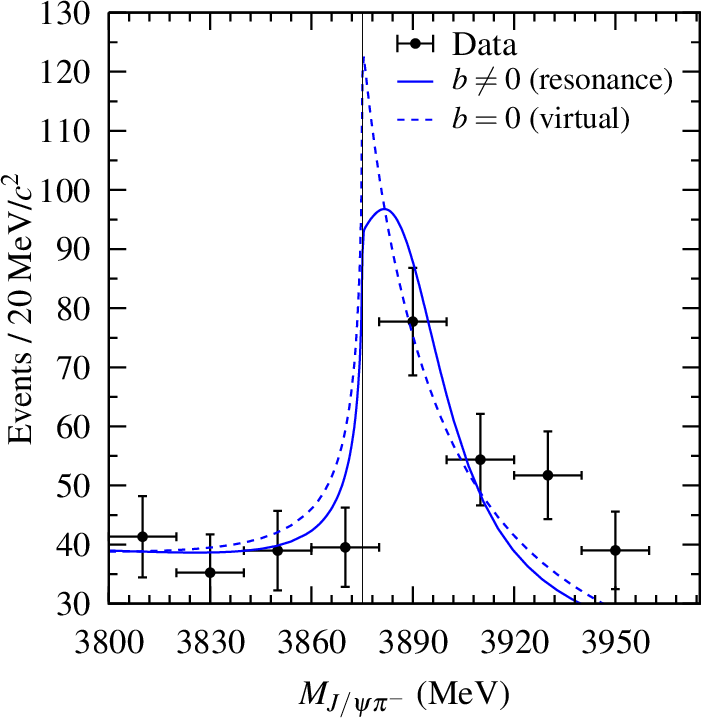}\hspace{2mm}
\includegraphics[width=0.33\textwidth,keepaspectratio]{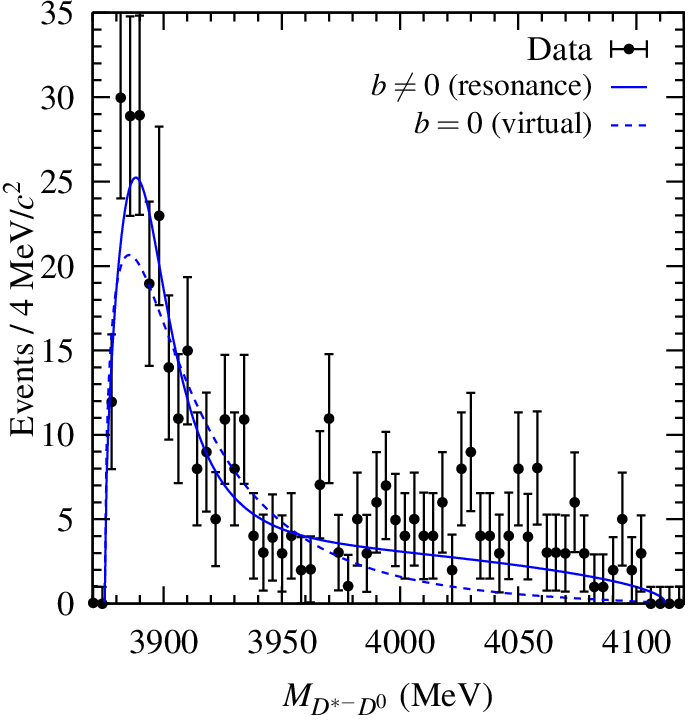}
\caption{Invariant mass distributions for $J/\psi \pi^-$ in the decay $Y(4260)
\to J/\psi \pi\pi$ (left panels) \cite{Ablikim:2013mio} and for
$\bar{D}^{\ast-}D^0$ in the decay $Y(4260) \to \bar{D}^\ast D \pi$  (right
panels) \cite{Ablikim:2015swa}. The top panels show the results for the fit
$b\neq 0$, $\Lambda_2=0.5\ \text{GeV}$. The inner and lighter error bands
reflect the statistical uncertainties, while the outer and darker bands  include
also the systematic ones. In the bottom panels, the two fits $b=0$ and $b\neq 0$
are compared (without error bands) for the case $\Lambda_2=0.5\ \text{GeV}$. In
the $J/\psi \pi^-$ spectrum, the $\bar{D}^\ast D$ threshold is marked with a
vertical black line.\label{fig:BothExpComparison}}
\end{figure*}

In principle, the double-$D$-tag technique ensures that all the
$\bar{D}^\ast D$ spectrum events in Ref.~\cite{Ablikim:2015swa} contain a $\bar{D}^\ast D$ pair, so there is no background due to wrong identification of the final
state. There could be, however, contributions to the spectrum from higher waves
other than the $S$-wave. In any case, an inspection of
Fig.~\ref{fig:BothExpComparison} shows that the tail of the spectrum is small,
and we set $\mathcal{B}_2=0$. We shall come to this point later on. For the
$J/\psi \pi$ spectrum, $\mathcal{B}_1$ is parameterized with a symmetric smooth
threshold function as used in the experimental work of
Ref.~\cite{Ablikim:2013mio}:
\begin{equation}\label{eq:JpsiBackground}
\mathcal{B}_1(s) = B_1 \left[(\sqrt{s}-m_{1-})(m_+-\sqrt{s})\right]^{d_1}~,
\end{equation}
with $m_{1-}=m_{J/\psi}+m_\pi$ and $m_+=m_Y-m_\pi$,
{\it i.e.}, the limits of the available phase space for the reaction. The
parameters $B_1$ and $d_1$ are free.

\begin{table*}
\begin{tabular}{ccccccc}
$\Lambda_2\ (\text{GeV})$ & $C_{1Z}\ (\text{fm}^2)$ & $b\ (\text{fm}^3)$ & $\widetilde{C}\ (\text{fm}^2)$ & $\chi^2/\text{dof}$ & $R_\text{th}$ \\ \hline
$1.0$ & $          - 0.19 \pm 0.08 \pm 0.01$ & $ -2.0 \pm 0.7 \pm 0.4$ & $0.39 \pm 0.10 \pm 0.02$ & $1.02$ & $\hphantom{1}6.0 \pm 3.5 \pm 0.5$\\
$0.5$ & $\hphantom{+}0.01 \pm 0.21 \pm 0.03$ & $ -7.0 \pm 0.4 \pm 1.4$ & $0.64 \pm 0.16 \pm 0.02$ & $1.09$ & $\hphantom{1}6.5 \pm 3.6 \pm 0.2$\\
$1.0$ & $          - 0.27 \pm 0.08 \pm 0.07$ & $0$ (fixed)             & $0.34 \pm 0.14 \pm 0.01$ & $1.31$ & $           10.3 \pm 9.0 \pm 1.1$\\
$0.5$ & $          - 0.27 \pm 0.16 \pm 0.13$ & $0$ (fixed)             & $0.54 \pm 0.16 \pm 0.02$ & $1.36$ & $           10.9 \pm 9.0 \pm 2.5$\\ \hline
\end{tabular}
\caption{Parameters of the $T$-matrix obtained for the different fits performed in this work, together with the reduced $\chi^2$ and the ratio $R_\text{th}$ obtained. The first (second) error is statistical (systematic). The pole position found for the $Z_c$ state in each case is given, in the same order as here, in Table~\ref{tab:poles}.\label{tab:params}}
\end{table*}
\begin{table}
\begin{tabular}{cclc}
$M_{Z_c}$ (MeV) & $\Gamma_{Z_c}/2$ (MeV) & Ref. & Final state \\ \hline
$3899 \pm 6$ & $23 \pm 11$ & \cite{Ablikim:2013mio} (BESIII) & $J/\psi\ \pi$ \\ % #### BESIII, PRL,110,252001 (2013), J/psi Pi, abstract, syst+stat added in quadrature
$3895 \pm 8$ & $32 \pm 18$ & \cite{Liu:2013dau} (Belle) & $J/\psi\ \pi$ \\ % #### BELLE, PRL,110,252002 (2013), J/psi Pi, abstract, syst+stat added in quadrature
$3886 \pm 5$ & $19 \pm \hphantom{1}5$ & \cite{Xiao:2013iha} (CLEO-c)& $J/\psi\ \pi$ \\ % #### CLEO-c data (Xiao, Dobbs, Tomaradze, Seth), PL,B727,366 (2013), JPsi Pi, abstract, syst+stat added in quadrature
$3884 \pm 5$ & $12 \pm \hphantom{1}6$ & \cite{Ablikim:2013xfr} (BESIII) & $\bar{D}^\ast D$\\ % #### BESIII, PRL,112,022001 (2014), D Dsbar,  abstract, syst+stat added in quadrature
$3882 \pm 3$ & $13 \pm \hphantom{1}5$ & \cite{Ablikim:2015swa} (BESIII) & $\bar{D}^\ast D$\\ \hline% #### BESIII, arXiv:1509.01398v1 (2015), D Dsbar, abstract, syst+stat added in quadrature
$3894 \pm 6 \pm 1$ & $30 \pm            12 \pm 6$ & $\Lambda=1.0$ GeV & $J/\psi\ \pi$, $\bar{D}^\ast D$\\
$3886 \pm 4 \pm 1$ & $22 \pm \hphantom{1}6 \pm 4$ & $\Lambda=0.5$ GeV & $J/\psi\ \pi$, $\bar{D}^\ast D$\\ \hline

$3831 \pm 26^{+\hphantom{1}7}_{-28}$            & virtual state & $\Lambda=1.0$ GeV & $J/\psi\ \pi$, $\bar{D}^\ast D$\\
$3844 \pm 19^{+           12}_{-21}$ & virtual state & $\Lambda=0.5$ GeV & $J/\psi\ \pi$, $\bar{D}^\ast D$\\ \hline
\end{tabular}
\caption{Mass and width of the $Z_c$ resonance reported in various experiments
and in this work. The first five rows show the values obtained in different experimental analyses
(statistical and systematical errors have been added in quadratures). The last
four rows correspond to the determinations from the different fits carried out
in this work, in the same order as shown in Table~\ref{tab:params}. In the
latter cases, the first (second) error is statistical (systematic). In the last two rows, corresponding to the case of a virtual state, we do not consider the small imaginary part ($\simeq 8\ \text{MeV}$) of the pole.
\label{tab:poles}}
\end{table}

%\section{Results}

We have three free parameters directly related to our $T$-matrix ($C_{1Z}$,
$\widetilde{C}$, and $b$), and six ($B_1$, $d_1$, $\alpha$, $\beta$ and
$K_{1,2}$) related to the background and the overall normalization.
These nine free parameters are adjusted to reproduce the data of
Refs.~\cite{Ablikim:2013mio,Ablikim:2015swa} (a total of 104 data points). In
this work, two errors are given.
The first error is statistical and it is computed from the hessian matrix of the
$\chi^2$ merit function. The second error is systematic, and to estimate it we
have considered two different uncertainty sources. First, we have varied the
$J/\psi \pi$ background function [Eq.~\eqref{eq:JpsiBackground}] and used other
smooth functions.
The second source of uncertainties is related to the tail of the $\bar{D}^\ast
D$ spectrum, and it is estimated as follows. The central value of the parameters is computed by
fitting this spectrum up to $\sqrt{s} = 4025\ \text{MeV}$. Then, we vary this
limit between $\sqrt{s} = 3975\ \text{MeV}$ and $m_+$ (the maximum allowed
invariant mass), and repeat the fit. In all cases, we find statistically
acceptable fits and the difference between the new fitted parameters and the
central ones is used to determine the systematic error. The same method is
applied to estimate the systematic error of our predictions for the spectra and
the mass and width of the $Z_c$ state, to be presented below.

We perform four different fits, corresponding to the two cases of keeping the
parameter $b$, which controls the energy dependence of the $\bar D^* D$
potential, free or set to zero, and for each of these, we choose $\Lambda_2$ to
be $0.5$ or $1\ \text{GeV}$ \cite{HQSSFormalism}. Results from the four fits are
compiled in Table~\ref{tab:params}, where only the parameters that are directly
related to our $T$-matrix are shown. One first notes that the reduced $\chi^2$
is very close to unity in all four cases. Indeed, the description of the
experimental spectra is very good in all cases, as can be seen in the top panels
of Fig.~\ref{fig:BothExpComparison}, where the results from one of the fits ($b$
free and $\Lambda_2=0.5\ \text{GeV}$) are shown and confronted with the data. In
particular, the effect of the $Z_c$ is nicely reproduced in the $\bar{D}^\ast D$
spectrum above threshold and in the $J/\psi \pi$ spectrum around the
$\bar{D}^\ast D$ threshold. Its reflection can also be appreciated in the
$J/\psi \pi$ distribution around $3.5\ \text{GeV}$. The other fits lead to
results similar to those shown in Fig.~\ref{fig:BothExpComparison}. The largest
differences can be found in the $\bar{D}^\ast D$ spectrum between the $b
\neq 0$ and $b=0$ cases, which are compared for $\Lambda_2 = 0.5\ \text{GeV}$ in
the bottom right panel of the same figure. In any case, we see that we
are able to simultaneously reproduce the two available BESIII data sets related to the
$Z^\pm_c(3900)/Z^\pm_c(3885)$ state with a single structure for the very first
time.
% and we find pole positions in fair agreement with the experimental analyses
% for this resonances.

\begin{figure}
\includegraphics[width=0.37\textwidth,keepaspectratio]{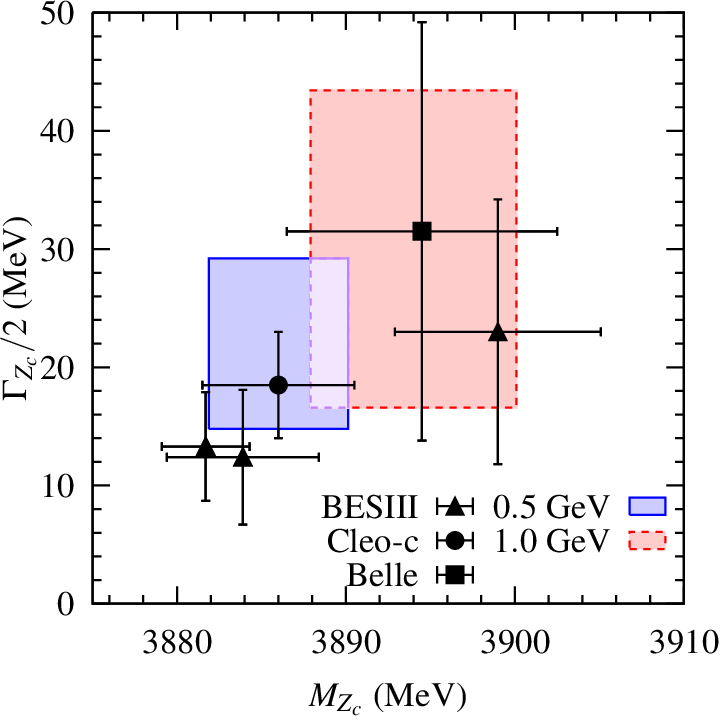}
\caption{Comparision of the $Z_c$ resonance pole positions determined in this
work for two values of the cutoff $\Lambda_2$ with the experimental
determinations of Refs.~\cite{Ablikim:2013mio, Ablikim:2013xfr, Ablikim:2015swa, Liu:2013dau, Xiao:2013iha}. The shaded areas take into account our statistical and systematic uncertainties (added in quadratures). The numerical values are shown in Table~\ref{tab:poles}.\label{fig:Poles}}
\end{figure}

Since we have a good description of the data where the $Z_c$ peak is seen, we
next study the pole structure of the $T$-matrix. Poles can be found in different Riemann sheets of the $T$-matrix, which are reached through analytical continuation of the $G$ functions in Eqs.~\eqref{eq:G1} and \eqref{eq:G2}. The $(\eta_1 \eta_2)$ Riemann sheet is defined with the following replacements:
\begin{align}
G_1(E) & \rightarrow G_1(E) + \eta_1 i \frac{q_1(E)}{4\pi E}~, \label{eq:G1_2rs}\\
G_2(E) & \rightarrow G_2(E) + \eta_2 i \frac{q_2(E)}{4\pi (m_D + m_D^\ast)}~.\label{eq:G2_2rs}
\end{align}
In this way, the physical sheet would be denoted as $(00)$.

We define the mass and the width of the $Z_c$ from its pole position, $\sqrt{s} = M_{Z_c} - i
\Gamma_{Z_c}/2$. For the case $b \neq 0$, we find poles on the $(11)$ Riemann
sheet, which is connected to the physical one above the $D\bar{D}^\ast$
threshold, at energies shown in Table \ref{tab:poles}. The real part of these
energies is clearly above threshold, so they correspond to a resonance, which
really (physically) exists as an unstable particle. In Fig.~\ref{fig:Poles}, we
compare the pole position obtained in this work for the $Z_c$ resonance with the
experimental determinations of Refs.~\cite{Ablikim:2013mio, Liu:2013dau,
Xiao:2013iha, Ablikim:2013xfr, Ablikim:2015swa}. Such comparisons are also
displayed in Table~\ref{tab:poles}. There is a good agreement within errors, and
the small differences can be traced back to the fact that these experimental
analyses used a Breit-Wigner parametrization which is not good around a
strongly-coupled threshold.
% Hence, we have shown that we are able to simultaneously reproduce the two
% available sets of BESIII experimental data related to the
% $Z^\pm_c(3900)/Z^\pm_c(3885)$ for the first time in the literature, and we
% find pole positions in fair agreement with the experimental analyses for this
% resonances.

For the case $b=0$, however, the situation is quite different. While the
description of the experimental data is still quite good with $\chi^2/\text{d.o.f.}\in
[1.3,1.4]$, the pole in this case is located below threshold, with a small
imaginary part (around $8\ \text{MeV}$), and in the $(01)$ Riemann sheet.
% This
% Riemann sheet is not continuously connected to the physical one, and hence,
% this pole cannot be associated to a physical state. However, the pole on this
% sheet produces a sizeable enhancement at $D\bar{D}^\ast$ threshold on the
% amplitude, as seen in Fig.~\ref{fig:BothExpComparison} (left bottom panel).
If the $J/\psi
\pi$ channel were now switched off ($\widetilde{C}=0$), this pole would move
into the real axis in the unphysical Riemann sheet of the elastic amplitude
$T_{22}$. In this sense, the obtained pole does not qualify as a resonance, and
we see it as a virtual or anti-bound $D\bar D^*$ state. It does not correspond
to a particle in the sense that its wave function, unlike that of a
bound state, is not localized. However, it produces observable effects at the
$D\bar{D}^\ast$ threshold similar to those produced by a near threshold resonance or bound state.\footnote{For example, in the triplet ${^3}S_1- {^3}D_1$
nucleon-nucleon waves there appears the deuteron, a truly bound state, with real existence (one can prepare a target
or a beam made up of this particle), while in the singlet $^1S_0$ wave there is
a virtual state, which has not real existence in this sense.} Indeed, scattering experiments alone, in principle, cannot distinguish between virtual and bound states, but the difference is not a purely academic one since they can
produce different line shapes in inelastic open channels~\cite{Hanhart:2007yq}.
The line shapes of a virtual state and a near-threshold resonance are different since the former is
peaked exactly at the threshold while the latter, in principle, is above. This
can be seen in the left bottom panel of Fig.~\ref{fig:BothExpComparison} where
the $J\psi \pi^-$ spectrum for the two fits $b=0$ and $b\neq 0$ are shown (for
the case $\Lambda_2 = 0.5\ \text{GeV}$). Although the two curves are different,
each one would approximately lie within the error band of the other. Clearly,
very precise data with a good energy resolution and small bin size are necessary
to distinguish among them.

Without taking sides, and given that both natures for the $Z_c$ structure
(resonance or virtual state) arise in fits of good quality, it must be stated
that the experimental information available at this time cannot fully
discriminate between both scenarios and, hence, claims about the $Z_c$ structure
should be made with caution. Nevertheless, the resonance scenario seems to be
statistically slightly preferred.
It is also clear that more experimental information is needed to elaborate on
the nature of $Z_c$. In particular, the spectrum of $J/\psi \pi$ with narrower
bins would be highly desirable to have a good resolution on its line shape. If
it is finally shown to be a virtual state, then it cannot be a tetraquark, since
it does not correspond to a normal particle, and it can only have a hadronic
molecular nature, in the sense that it appears only because of the $D\bar D^*$
interaction.

Summarizing, we have studied the two decays ($Y(4260) \to J/\psi \pi^+\pi^-,
D^{\ast-}D^0 \pi^+$) in which the $Z^{\pm}_c$ resonant-like structure is seen.
We have presented the first simultaneous study of the invariant mass
distributions of the $J/\psi\pi$ and $\bar{D}^\ast D$ channels with fully
unitarized amplitudes. We find that these data sets are well reproduced in two different scenarios. In
the first one, in which there is an energy dependence in the $\bar{D}^\ast D \to
\bar{D}^\ast D$ potential, the $Z_c$ appears as a dynamically generated
$\bar{D}^\ast D$ resonance.
In the second one, however, when the aforementioned energy dependence is not
allowed, it appears as a virtual state, with the pole located
below the $D\bar D^*$ threshold. In any case, it is demonstrated that both data
sets can be reproduced with only one $Z_c$ state, so that the two experimentally
observed structures $Z^\pm_c(3900)$ and $Z_c^\pm(3885)$, in different channels,
are proven to correspond to the same state. Moreover, both fits do not
allow a $\bar D^* D$ bound state solution.\footnote{We use the term ``bound
state'' loosely here as if all inelastic channels including the $J/\psi\pi$ are
neglected.} Since the virtual state can only be of hadronic-molecule type, it is really important to discriminate between these two scenarios. For that purpose, one needs a very precise measurement of the line shapes around, in particular slightly above, the $D\bar D^*$ threshold.
Such a measurement is foreseen when more data are collected at BESIII.

\section*{Acknowledgments}

We thank Christoph Hanhart for a careful reading of the manuscript and for his
very useful and constructive comments. F.-K.~G. is grateful to IFIC for the
hospitality during his visit, and is partially supported by the Thousand Talents Plan for Young
Professionals. C.~H.-D. thanks the support of the JAE-CSIC Program.
M.~A.
acknowledges financial support from the ``Juan de la Cierva'' program
(27-13-463B-731) from the Spanish MINECO. This work is supported in part by the
DFG and the NSFC through funds provided to the Sino-German CRC 110 ``Symmetries
and the Emergence of Structure in QCD'' (NSFC Grant No. 11261130311), by the
Spanish MINECO and European FEDER funds under the contract FIS2011-28853-C02-02
and FIS2014-51948-C2-1-P, FIS2014-57026-REDT and SEV-2014-0398, by Generalitat
Valenciana under contract PROMETEOII/2014/0068, by the EU HadronPhysics3 project
with grant agreement no.
283286.

\end{document}